\documentclass{elsart}

\usepackage{graphics}
\usepackage{epsfig}
\usepackage{amsfonts}
\usepackage{amssymb}
\usepackage{cite}
\usepackage{here}

\begin{document}

\begin{frontmatter}
\title{Polarisation of the $\omega$ meson in the $pd \to
^{\,3\!\!}\textrm{He}\,\omega$ \\ reaction at 1360 and 1450~MeV}

\author[UU]{K.~Sch\"onning},
\author[Stock]{Chr.~Bargholtz},
\author[Tuebingen]{M.~Bashkanov},
\author[SINS]{M.~Ber{\l}owski},
\author[Dubna]{D.~Bogoslawsky},
\author[UU]{H.~Cal\'en},
\author[Tuebingen]{H.~Clement},
\author[Hamburg]{L.~Demir\"ors},
\author[TSL]{C.~Ekstr\"om},
\author[UU]{K.~Fransson},
\author[Stock]{L.~Ger\'en},
\author[UU]{L.~Gustafsson},
\author[UU]{B.~H\"oistad\corauthref{cor1}},
\corauth[cor1]{Corresponding author.\\Bo H\"oistad, Physics and Astronomy Department, Uppsala University,\\Box 535, 751 21 Uppsala, Sweden} \ead{bo.hoistad@tsl.uu.se}
\author[Dubna]{G.~Ivanov},
\author[UU]{M.~Jacewicz},
\author[Dubna]{E.~Jiganov},
\author[UU]{T.~Johansson},
\author[UU]{S.~Keleta},
\author[Tuebingen]{O.~Khakimova},
\author[UU]{I.~Koch},
\author[Tuebingen]{F.~Kren},
\author[UU]{S.~Kullander},
\author[UU]{A.~Kup\'s\'c},
\author[Novosibirsk]{A.~Kuzmin},
\author[Stock]{K.~Lindberg},
\author[UU]{P.~Marciniewski},
\author[Dubna]{B.~Morosov},
\author[Juelich]{W.~Oelert},
\author[Hamburg]{C.~Pauly},
\author[UU]{H.~Pettersson},
\author[Dubna]{Y.~Petukhov},
\author[Dubna]{A.~Povtorejko},
\author[UU]{R.J.M.Y.~Ruber},
\author[Hamburg]{W.~Scobel},
\author[MEPI]{R.~Shafigullin},
\author[Novosibirsk]{B.~Shwartz},
\author[Tuebingen]{T.~Skorodko},
\author[ITEP]{V.~Sopov},
\author[SINS]{J.~Stepaniak},
\author[Stock]{P.-E.~Tegn\'er},
\author[UU]{P.~Th\"orngren Engblom},
\author[Dubna]{V.~Tikhomirov},
\author[WU]{A.~Turowiecki},
\author[Tuebingen]{G.J.~Wagner},
\author[UCL]{C.~Wilkin},
\author[Juelich]{M.~Wolke},
\author[SINS]{J.~Zabierowski},
\author[Stock]{I.~Zartova},
\author[UU]{J.~Z{\l}oma\'nczuk}

\collab{CELSIUS/WASA Collaboration}

\address[UU]{Uppsala University, Uppsala, Sweden}
\address[Stock]{Stockholm University, Stockholm, Sweden}
\address[Tuebingen]{Physikalisches Institut der Universit\"at T\"ubingen,
T\"ubingen, Germany}
\address[SINS]{Soltan Institute of Nuclear Studies, Warsaw and Lodz, Poland}
\address[Dubna]{Joint Institute for Nuclear Research, Dubna, Russia}
\address[Hamburg]{Institut f\"ur Experimentalphysik der Universit\"at Hamburg,
Hamburg, Germany}
\address[TSL]{The Svedberg Laboratory, Uppsala, Sweden}
\address[Novosibirsk]{Budker Institute of Nuclear Physics, Novosibirsk, Russia}
\address[Juelich]{Institut f\"ur Kernphysik, Forschungszentrum J\"ulich,
J\"ulich, Germany}
\address[MEPI]{Moscow Engineering Physics Institute, Moscow, Russia}
\address[ITEP]{Institute of Theoretical and Experimental Physics,
Moscow, Russia}
%\address[Osaka]{Research Centre for Nuclear Physics, Osaka, Japan}
\address[WU]{Institute of Experimental Physics of Warsaw University, Warsaw, Poland}
\address[UCL]{Physics and Astronomy Department, UCL, London, United Kingdom}

\newpage

\begin{abstract}
The tensor polarisation of $\omega$ mesons produced in the $pd \to
^{\,3\!\!}\textrm{He}\,\omega$ reaction has been studied at two
energies near threshold. The $^3$He nuclei were detected in coincidence
with the $\pi^0\pi^+\pi^-$ or $\pi^0\gamma$ decay products of the
$\omega$. In contrast to the case of $\phi$-meson production, the
$\omega$ mesons are found to be unpolarised. This brings into question
the applicability of the Okubo-Zweig-Iizuka rule when comparing
the production of vector mesons in low energy hadronic reactions.
\end{abstract}

\begin{keyword}
$\omega$ meson production; $\omega$ polarisation; OZI rule%
\PACS 25.40.Ve %Other reactions above meson production thresholds (energies > 400 MeV)%
\sep 14.40.Cs %Other mesons with S=C=0, mass < 2.5 GeV%
\sep 13.88.+e %Polarization in interactions and scattering %
\end{keyword}
\end{frontmatter}
%
%%%%%%%%%%%%%%%%%%%%%%%%%%%%%%%%%%%%%%%%%%%%%%%%%%%%%%%%%%%%%%%%%
%
\newpage

The production of the light isoscalar vector mesons $\phi $ and
$\omega $ in various nuclear reactions involving non-strange
particles are often compared within the framework of the
Okubo-Zweig-Iizuka rule~\cite{OZI}. This rule suggests that processes with broken quark lins are suppressed, and therfore the cross section ratio between $\phi $ and
$\omega $ production  is mainly due
to small deviations from ideal mixing of these mesons at the quark
level. The ratio of the squares of the production amplitudes for
the two mesons, for any hadronic reaction measured under similar
kinematic conditions, should be of the order of
$R_{\phi/\omega}\approx R_{\mathrm{OZI}}=4.2\times
10^{-3}$~\cite{Lipkin}. The validity of this estimate has been
tested for the $pd\rightarrow\,^{3}\textrm{He}\,\omega/\phi$
reaction near threshold, where it was found that
$R_{\phi/\omega}\approx 20\times
R_{\mathrm{OZI}}$~\cite{Wurzinger95,Wurzinger96,MOMO}. This
deviation is over a factor of two greater than that found, for
example, in the case of production in nucleon-nucleon collisions
near threshold~\cite{Hartmann,Maeda}. It is thus unclear to what extent the OZI approach is applicable for this reaction, and further experimental input would be valuable. 

In the MOMO study of the $pd\rightarrow\,^{3}\textrm{He}\,\phi$
reaction~\cite{MOMO}, the $K^+$ and $K^-$ coming from the decay of
the $\phi$ were measured in coincidence with the $^3$He ejectile.
Now the angular distribution of the $K^{+}K^{-}$ relative momentum
in the rest frame of the $\phi$-meson is sensitive to the tensor
polarisation (spin alignment) of the spin-one meson. The
surprising result from the MOMO experiment is that near threshold
the $\phi$ are produced almost purely in the magnetic sub-state
with $m=0$ along the beam direction~\cite{MOMO}. In the light of the OZI consideration in comparing the cross sections of the $\omega$ and $\phi$ production it should also be interesting to compare the polarisation of these mesons produced in the $pd\rightarrow\,^{3}\textrm{He}\,\omega/\phi$ reactions, since any difference in the $\omega/\phi$ polarisation is not expected to depend on the details of the quark mixing but rather on the reaction mechanism.  

The only two significant decay channels of the $\omega$ meson are
$\omega \rightarrow\pi^{0}\pi^{+}\pi^{-}$ and $\omega
\rightarrow\pi^{0}\gamma$, with branching ratios of 89.1\% and
8.7\%, respectively~\cite{PDG}. The angular distributions of both
decays reflect the spin alignment of the $\omega$. By measuring
both these channels, we obtained two different measurements  for the $\omega$ polarisation in the
$pd\rightarrow\,^{3}\textrm{He}\,\omega$ reaction. 

The measurements of the $\omega$ polarisation were carried out at the The Svedberg Laboratory in
Uppsala, Sweden, using the WASA detector~\cite{Zabierowski,WASA}
which, until June 2005, was an integral part of the CELSIUS
storage ring. The experiments were done at $T_{p}=1360$ and
1450~MeV, corresponding to excess energies of 17 and 63~MeV with
respect to the nominal $^3\textrm{He}\,\omega$ threshold. The
circulating proton beam was incident on deuterium pellet
targets~\cite{Ekstrom, Nordhage}. The $^{3}$He ejectiles were
measured in the WASA forward detector (FD), which covered
laboratory polar angles from $3^{\circ}$ to $18^{\circ}$. This
corresponds to 95\% of the $^{3}$He phase space for $\omega$
production at 1450~MeV and 78\% at 1360~MeV. The majority of the
lost events are those where the $^{3}$He are emitted at small
laboratory angles such that they escape detection down the beam
pipe. The corresponding angular acceptance of the $\omega$ mesons
covers, in the CM system, the intervals $22^{\circ}-158^{\circ}$
at 1450 MeV and $46^{\circ}-134^{\circ}$ degrees at 1360 MeV.

\begin{figure}[hbt]
\begin{center}
\includegraphics[width=0.8\textwidth]{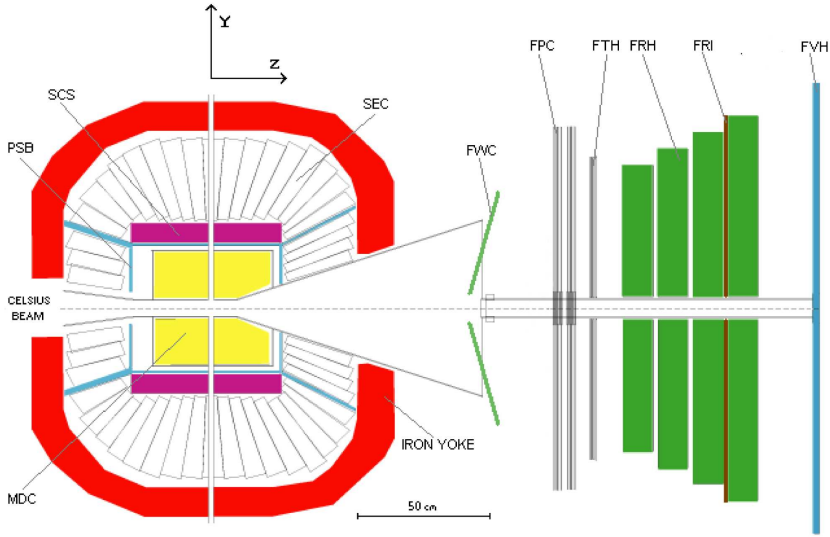}
\end{center}
\caption{Side view of the CELSIUS/WASA detector setup. The CELSIUS
beam pipe runs horizontally and the target pellets are injected
downwards through the vertical pipe.} \label{fig:wasa4pi}
\end{figure}

The forward detector consists of sector-like scintillations
detectors forming a window counter (FWC) for triggering, a
hodoscope (FTH) for triggering and off-line particle
identification, a range hodoscope (FRH) for energy measurements,
particle identification and triggering, and a veto hodoscope (FVH)
for triggering. A proportional chamber (FPC) for precise angular
information is also part of the forward detector. Mesons and their
decay products are mainly measured in the central detector (CD)
that consists of the Plastic Scintillating Barrel (PSB), the Mini
Drift Chamber (MDC), the Super Conducting Solenoid (SCS), and the CsI equipped Scintillating
Electromagnetic Calorimeter (SEC). The SEC, which measures the
angles and energies of photons arising from meson decay, covers
polar angles from $20^{\circ}$ to $169^{\circ}$. A schematic
overview of the setup is shown in Fig.~\ref{fig:wasa4pi}.

%\vspace{5mm}

The hardware $^3$He trigger selected events where there was a hit
with a high energy deposit in the FWC as well as a hit in either
the FTH (March 2005 run) or the FRH (May 2005 run) in the same
$\phi$ angle sector.

The $^3$He was identified in the FD using the $\Delta E\!-\!E$
method, as described in Ref.~\cite{karin1}. Here the light output
from the detector layer where the particle stopped was compared
with that from the preceding layers. The $\chi^2$ of a particular
particle hypothesis was then calculated by comparing the measured
energy deposits in all detector layers traversed to those expected
for that particle. Particle hypotheses giving a $\chi^2$ larger
than a maximum value, chosen to reduce background without losing
good events, were rejected~\cite{jozef}.

\begin{figure}[hbt]
\begin{center}
\includegraphics[width=0.9\textwidth]{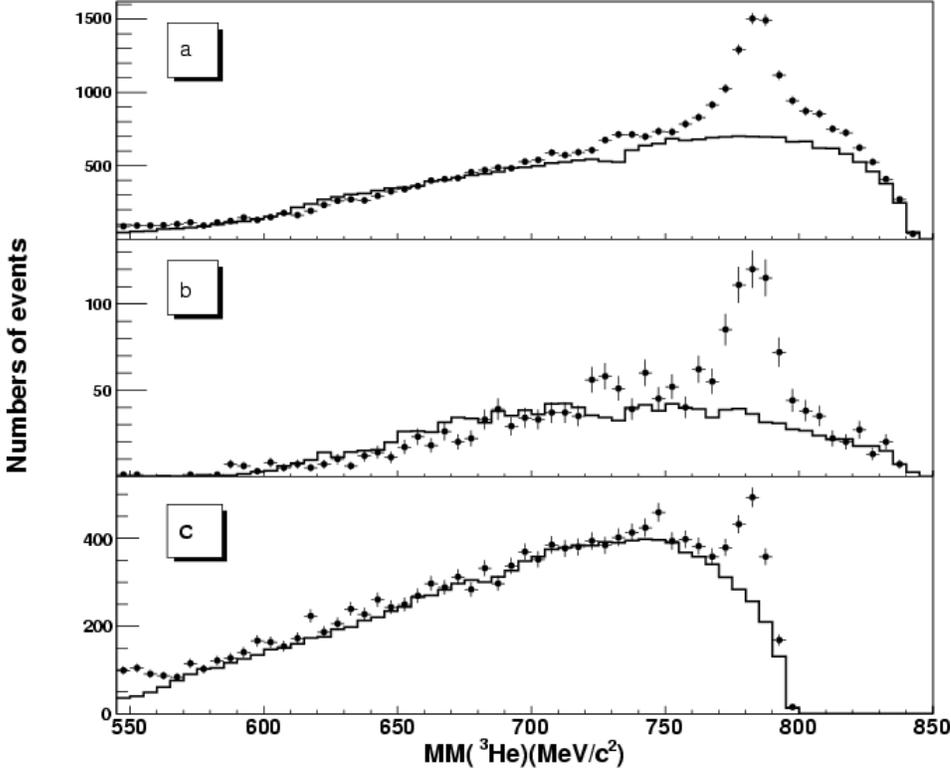}
\end{center}
\caption{a) The points show the missing mass distribution from all
1450~MeV data that fulfill the criteria optimised for the
selection of $pd\rightarrow \,^{3}\textrm{He}\,(\omega\rightarrow
\pi^0\pi^+\pi^-)$ events, as explained in the text. The histogram
represents a Monte Carlo simulation of the $pd\rightarrow
\,^{3}\textrm{He}\,\pi^0\pi^+\pi^-$ reaction, assuming phase-space
production. b) The data show the missing mass distribution for the
$pd\rightarrow\,^{3}\textrm{He}\,(\omega\rightarrow\pi^0\gamma)$
event selection, while the histogram represents a phase-space
simulation of the $pd\rightarrow \,^{3}\textrm{He}\,\pi^0\pi^0$
reaction. c) The same as Panel-a, but for the 1360~MeV data.}
\label{fig:mm1450}
\end{figure}

The main focus of the present work is on the
$\omega\rightarrow\pi^{0}\pi^{+}\pi^{-}$ decay channel, where the
large branching ratio (89.1\%) gives the highest statistics. To
select this channel we require one $^{3}$He with a well defined
energy and angle in the FD and at least two photons in the SEC. In
addition, one photon pair must have an invariant mass close to
that of the $\pi^{0}$. The missing mass of the $^{3}$He$\,\pi^{0}$
system must be larger than 250~MeV/c$^{2}$, \emph{i.e}, twice the
pion mass folded with the experimental resolution, in order to
select the events with two additional pions. Two or more hits are
required in the PSB. Finally, we require at least one track in the
MDC coming from the overlap region between the pellet target and
the proton beam. The missing mass of the $^{3}$He is shown in
Fig.~\ref{fig:mm1450}a for all events fulfilling the above
criteria. \vspace{5mm}

The selection requirements lead to an overall acceptance of 14\%
at both beam energies. In addition to the losses at small angles
in the beam pipe, there are losses from the $^{3}$He that undergo
nuclear interactions before depositing all their energy. Moreover
photons from $\pi^{0}$ decay can escape detection in the CD and,
finally, there is the limited MDC efficiency ($\approx 50\%$).
About 10\% (30\%) of the events at 1450 (1360)~MeV are produced
outside the pellet target (mainly in beam-rest gas interactions)
and are therefore rejected.

For an $\omega$ meson decaying into $\pi^{0}\pi^{+}\pi^{-}$, the
spin direction can be specified with respect to an axis directed
along the normal to the decay plane. This direction is given by
the vector product of the momenta of the $\pi^{0}$ and one of the
charged pions in the rest frame of the $\omega$ meson. For this
purpose, the $\pi^{0}$ was reconstructed from the decay photons,
and the charged pion from the precise angular determination in the
MDC combined with the information from the $^{3}$He and the
$\pi^{0}$.

The polarisation can be measured by studying the dependence of the
cross section on the angle $\beta$ between the normal and some
quantisation axis in the \textit{Gottfried-Jackson}
frame~\cite{gottfriedjackson}, \textit{i.e.}, the rest frame of
the $\omega$. For the Jackson angle the quantisation axis is taken
to be along the direction of the proton beam.

We are interested in the elements of the spin-density matrix
$\rho_{mm^{\prime}}$ that represent the tensor polarisation
(alignment) of the $\omega$. With an unpolarised beam and target,
there is one independent term
$\rho_{11}=\rho_{1-1}=\frac{1}{2}(1-\rho_{00})$ that can be
measured. The dependence of the differential cross section on
$\beta$ is of the form:
\begin{equation}
\frac{\mbox{\rm d}\sigma (\omega \rightarrow
\pi^{0}\pi^{+}\pi^{-})}{\mbox{\rm d}\cos \beta}\propto
(1-\rho_{00})+(3\rho_{00}-1)\cos^{2}\beta \,. \label{eq:1}
\end{equation}%
If the $\omega$ mesons are unpolarised, one has that $\rho
_{00}=\rho_{11}=\rho_{1-1}=\frac{1}{3}$ and thus an isotropic
angular distribution, while the maximum polarisation occurs when $\rho
_{00}=1$ and thus the distribution has a pure $\cos^2$ dependence. 

In order to obtain the differential cross section as a function of
$\cos^{2}\beta$, all events fulfilling the selection criteria were
divided into eight regions of $|\cos \beta |$. In view of the
limited statistics, no account was here taken of the $\omega$
direction in the CM system. In each region of $|\cos \beta |$ the
missing mass of the $^{3}$He (MM($^{3}$He)) was plotted. The
$\omega$ candidates show up in a peak near the nominal mass at
782.6~MeV/c$^{2}$, as clearly seen in the event distribution shown
in Fig.~\ref{fig:mm1450}a. The background under the $\omega$ peak
was estimated in two ways, either by taking a phase-space Monte
Carlo simulation of $pd\rightarrow \,^{3}\textrm{He}\,\pi^{0}\pi
^{+}\pi^{-}$ or by fitting the data to a gaussian peak on a
polynomial background. The difference in the numbers of $\omega$
obtained in the two ways is between 2\% and 15\%.

\begin{figure}[hbt]
\begin{center}
\includegraphics[width=0.8\textwidth]{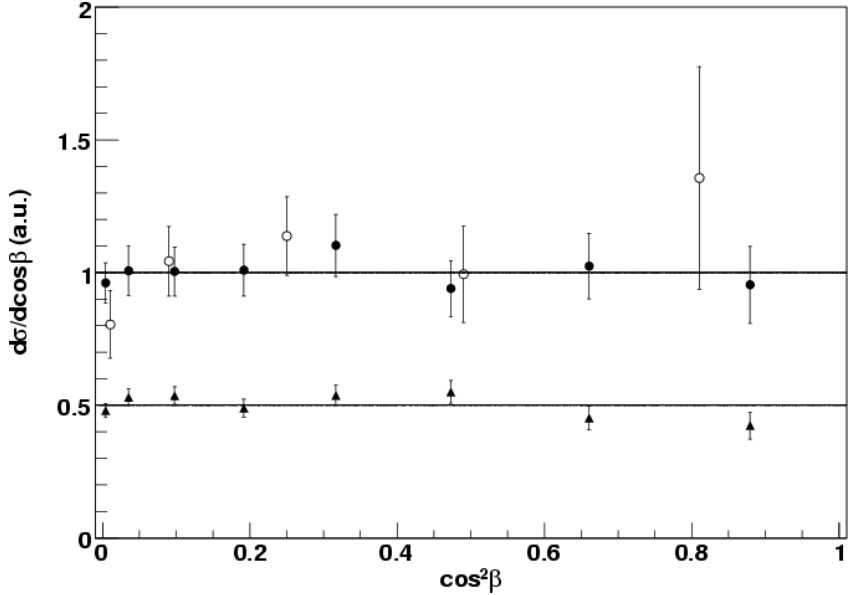}
\end{center}
\caption{The filled circles represent the differential cross
section for $pd\rightarrow \,^{3}\textrm{He}\,(\omega \rightarrow
\pi^0\pi^+\pi^-)$ at 1450~MeV as a function of $\cos^2\beta$, where $%
\beta$ is the angle between the normal to the $\omega$ decay plane
and the proton beam direction. The data are arbitrarily normalised
so that their average is unity, as indicated by the horizontal
line, and the error bars are purely statistical. The non-filled
circles show the corresponding cross section for the
$pd\rightarrow \,^{3}\textrm{He}\,(\omega \rightarrow\pi^0\gamma)$
channel, where $\beta^{\prime}$ now is the angle between the
$\pi^0$ and the incoming proton. Both cross sections have been
corrected for acceptance and are normalised using the identical
overall factor. The triangles represent the $\cos^2\beta$
distribution for $pd\rightarrow
\,^{3}\textrm{He}\,\pi^0\pi^+\pi^-$ for events with $700 <
MM(^{3}\textrm{He}) < 750$~MeV/c$^2$. The points were normalised
to unity but then shifted downwards by 0.5 to improve the
readability.} \label{fig:pol1450}
\end{figure}

In Fig.~\ref{fig:pol1450}, the angular dependence of the $\omega$
cross section at 1450\thinspace MeV is shown by the filled
circles. These have been normalised by an arbitrary factor to give
an average value of unity. Our data are clearly consistent with an
isotropic distribution. To investigate the situation further, the
same exercise was undertaken for events outside the peak region,
\textit{i.e.}, $700<MM(^{3}\textrm{He})<750$~MeV/c$^{2}$, where an
isotropic distribution is likely due to the many available states
for the multipion production as opposed to the $J^{P}=1^{-},T=0$
state of the $\omega$. After correcting for acceptance, the
corresponding points in Fig.~\ref{fig:pol1450} have been shifted
downwards by $0.5$ to improve the readability in the figure. The
statistics here are high and there is indeed little sign of any
angular dependence. This gives some confidence that our setup and
analysis does not introduce an artificial signal in the $\omega$
case.

A valuable consistency check can be obtained through the parallel
study of the $\omega \rightarrow \pi^{0}\gamma$ decay channel.
Although the low branching ratio of 8.7\% leads to poor
statistics, the signal-to-background ratio is better. This is due
to the fact that all final-state particles, \textit{i.e.}\ the
$^{3}$He and three photons, are measured with good acceptance.

In the event selection for this channel a $^{3}$He plus three
photons are demanded, where one photon pair has an invariant mass
close to that of the $\pi^{0},$ and the invariant mass of all
three photons should be larger than 600~MeV/c$^{2}$. The magnitude
of the missing mass in the $^{3}$He$\,3\gamma $ system must not
exceed 100~MeV/c$^{2}$ and the difference between the direction of
the missing momentum of the $^{3}$He and that of the
3$\gamma$-system may not be larger than $20^{\circ}$. Finally, we
apply the coplanarity cut on the laboratory azimuthal angles:
$160^{\circ}<|\phi_{\mathrm{lab}}(^{3}$He$)-\phi_{\mathrm{lab}}(3\gamma)|<200^{\circ}$.

The above conditions give an acceptance of 19\% (18\%) at 1450
(1360)~MeV and the data fulfilling these criteria are shown in
Fig.~\ref{fig:mm1450}b. The main background channel is
$pd\rightarrow \,^{3}\textrm{He}\,\pi^0\pi^0$ which, despite the
low acceptance of only 1.8\% for the given cuts, contributes
significantly due to the high production cross section.

The corresponding angle for the polarisation study in the $\omega
\rightarrow \pi^{0}\gamma$ channel is that of the $\pi^{0}$ or the
$\gamma$ in the $\omega$ rest frame with respect to the incoming
proton beam. Denoting this angle by $\beta^{\prime}$, the angular
distribution is expected to be of the form
\begin{equation}
\frac{\mbox{\rm d}\sigma (\omega \rightarrow
\pi^{0}\gamma)}{\mbox{\rm d} \cos \beta^{\prime}}\propto
(1+\rho_{00})-(3\rho_{00}-1)\cos ^{2}\beta^{\prime}\,.
\label{eq:1a}
\end{equation}

Since the statistics are smaller than for the three-pion decay,
the data were divided into five regions in $|\cos\beta^{\prime}|$.
The number of $\omega$ candidates in each region was also obtained
by plotting the missing mass of the $^{3}$He and subtracting the
background, both by using fitted Monte Carlo simulations of
$pd\rightarrow \,^{3}\textrm{He}\,\pi^{0}\pi^{0}$ and by fitting a
gaussian peak plus a polynomial. The differences in this case are
typically 2--25\%. After correcting for acceptance, branching
ratio and bin size, we normalise using the \textit{identical}
factor to that employed in the $\omega \rightarrow
\pi^{0}\pi^{+}\pi^{-}$ case. The good agreement in normalisation
between the two decay channels, seen in Fig.~\ref{fig:pol1450} at
1450~MeV, shows that the relative cut efficiencies and other
systematic effects are well understood. The data are also
consistent with isotropy, with a $\chi^{2}/\mathit{ndf}=1.02$,
though with much larger error bars than for the three-pion
channel.

The corresponding analysis at 1360~MeV is made much more difficult
by the relatively large width of the $\omega$. This becomes more
important as threshold is approached since the high mass tail of
the $\omega$ cannot then be produced and this leads to an
asymmetric peak. The background fitting is also more complicated
since the continuum ends under the $\omega$ peak and, furthermore,
the signal-to-background ratio is smaller than at 1450~MeV. It is
seen from Fig.~\ref{fig:mm1450}c that, within the interval
$750<MM(^{3}\textrm{He})<800$~MeV/c$^{2}$, the ratio is 1:5
compared to the 1:3 in the 1450~MeV case. In addition, more beam
time was taken at 1450 than at 1360~MeV so that there are both
higher systematic and statistical uncertainties at this energy.

\begin{figure}[tbp]
\begin{center}
\includegraphics[width=0.8\textwidth]{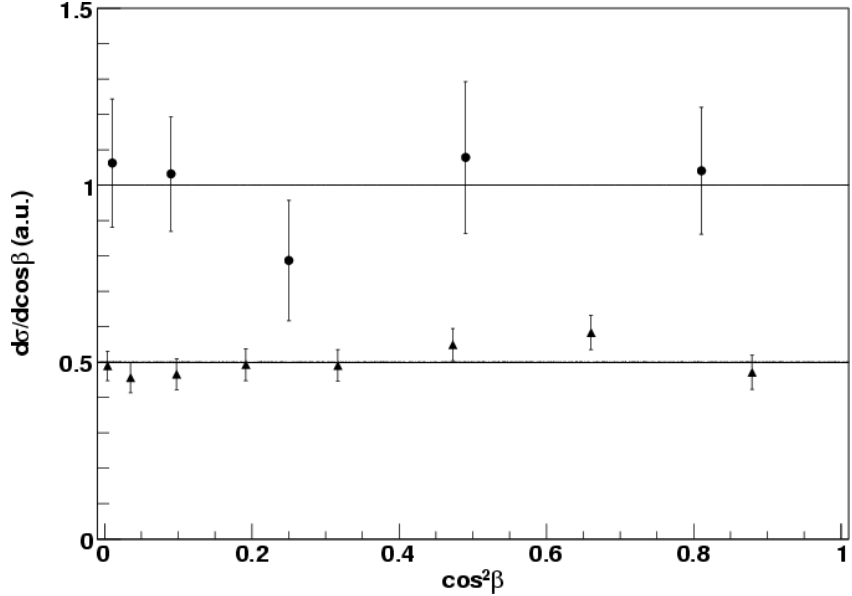}
\end{center}
\caption{The differential cross section for
$pd\rightarrow\,^{3}\textrm{He}\,(\omega \rightarrow
\pi^0\pi^+\pi^-)$ at 1360~MeV. As in Fig.~\ref{fig:pol1450}, the
data have been arbitrarily normalised to give an average of unity.
In addition to the statistical uncertainties shown, there are
uncertainties in the background subtraction of between 18\% and
30\%. The three-pion background from the region
$700<MM(^{3}\textrm{He})<750$~MeV/c$^2$ is also shown, normalised
to unity and shifted downwards by $0.5$.} \label{fig:pol1360}
\end{figure}

The statistics at 1360 MeV allow for a division of the $\omega
\rightarrow \pi^{0}\pi^{+}\pi^{-}$ data into only five regions of
$|\cos \beta |$ but, apart from this, the procedure for extracting
the differential cross section, shown as a function of $\cos
^{2}\beta$ in Fig.~\ref{fig:pol1360}, is exactly the same as that
at 1450~MeV. The systematic uncertainties from background
subtraction are between 18\% and 30\%. Within the much larger
error bars, the data are consistent with isotropy. This is also
the case for the three-pion background, selected in this case for
$700<MM(^{3}\textrm{He})<750$~MeV/c$^{2}$. Regarding the $\omega
\rightarrow \pi^{0}\gamma$ decay channel, the statistics at
1360~MeV are too poor to enable an investigation.

The angle between the $\pi^0$ from the $\omega \rightarrow
\pi^0\pi^+\pi^-$ decay and the incoming proton has also been
studied. In this case the distribution should be described by
Eq.~(\ref{eq:1a}) and so one effectively loses a factor of two in
sensitivity to the polarisation due to the larger constant factor
in the equation compared to Eq.~(\ref{eq:1}). The angular
distribution of the $\pi^0$ is consistent with isotropy at both
energies.

The angular dependence of the
$\omega\rightarrow\pi^{0}\pi^{+}\pi^{-}$ data shown at 1450 and
1360~MeV in Figs.~\ref{fig:pol1450} and \ref{fig:pol1360},
respectively, have been fitted by straight lines to extract the
values of $\rho_{00}$ on the basis of Eq.~(\ref{eq:1}). In this
way we obtain $\rho_{00}=0.33\pm 0.05$ at 1450~MeV and
$\rho_{00}=0.34\pm 0.10$ at 1360~MeV. The errors here are
statistical but it is clear that the deviation from an unpolarised
value of $\frac{1}{3}$ must be quite small. Some confirmation of
this is found at 1450~MeV through the study of the $\omega
\rightarrow \pi^{0}\gamma$ data, which gives $\rho _{00}=0.14\pm
0.14$. The uncertainty in this case is much larger than for the
three-pion channel, due mainly to the poorer statistics that
forced the reduction in the number of $|\cos\beta|$ intervals. It
should be noted though that the obtained deviation from the unpolarised
$\rho_{00}=\frac{1}{3}$ is here in the opposite direction
from that found at MOMO for the $\phi$ meson, having $\rho_{00}=0.82\pm0.05$ ~\cite{MOMO}.

In order to test whether the $\omega$ polarisation depends upon
its production angle $\theta_{\omega}^*$ in the overall CM system,
the $\omega \rightarrow \pi^0\pi^+\pi^-$ data at 1450~MeV were
divided into three sub-samples with respect to
$\cos\theta_{\omega}^*$. In all three regions the results were
consistent with $\omega$ being unpolarised. Specifically, for
\begin{eqnarray*}
\cos\theta_{\omega}^*\phantom{|} < -0.5\ \ \ &\Rightarrow&\ \ \ \rho_{00} =
0.29 \pm 0.08, \\
|\cos\theta_{\omega}^*| < \phantom{-}0.5\ \ \ &\Rightarrow&\ \ \ \rho_{00} =
0.37 \pm 0.06, \\
\cos\theta_{\omega}^*\phantom{|}> \phantom{-}0.5\ \ \ &\Rightarrow&\ \ \
\rho_{00} = 0.30 \pm 0.09.
\end{eqnarray*}

In this work we have shown that, within error bars, the $\omega$
produced in the $pd \rightarrow ^{\,3\!\!}\textrm{He}\,\omega$
reaction are unpolarised with respect to the incident proton beam
direction. However, it is also of interest to study the
polarisation in the helicity frame~\cite{gottfriedjackson}, where
the reference axis is provided by the direction of the $^3$He.
Unlike the Jackson angle distribution, this cross section must be
flat near threshold since only $s$-waves then contribute. Although
the sensitivity to the polarisation is small, it is reassuring
that the helicity distribution is completely consistent with
isotropy at both beam energies.

The contrast between our result for the $\omega$ polarisation and that of the $\phi$ in the $pd \rightarrow
^{\,3\!\!}\textrm{He}\,\omega/\phi$ reactions could hardly be more
striking since the MOMO collaboration reported a polarisation
along the proton beam direction corresponding to almost complete alignement ~\cite{MOMO}. 
Although the $\phi$ production was carried out slightly closer to threshold than the $\omega$ production,
it nevertheless suggests strongly that the reaction mechanism for the
production of the two mesons must differ significantly in their
details. An OZI inspired interpretation of the difference in the $\phi$ and
$\omega$ production near threshold therefore seems to fall short, both on the account of the new polarisation data as well as the previously reported cross section data. It is
therefore likely that the reactions are much more influenced
by nuclear and mesonic degrees of freedom rather than hadron properties at
the quark level.

It would be instructive if the $\omega$ and $\phi$ polarisations
could be compared in other low energy hadronic reactions. However,
it should be noted that conservation laws require that these
vector mesons must be completely polarised, with $\rho_{00}=0$
along the beam direction, when they are produced in
$pp\rightarrow{}pp\,\omega/\phi$ or
$pn\rightarrow{}d\,\omega/\phi$ at
threshold~\cite{DISTO,Hartmann,Maeda}. Hence any test here would
have to be carried out at higher energies.

In summary, from the study of both the $\omega \rightarrow
\pi^{0}\pi^{+}\pi^{-}$ and $\omega \rightarrow \pi^{0}\gamma$
decay channels in the $pd\rightarrow \,^{3}\textrm{He}\,\omega$
reaction, we have shown that the $\omega$ mesons are essentially
unpolarised near threshold. The relative cross sections obtained
using the two channels gives results that are consistent within
statistics. However, the absolute values of the cross section
depend upon the luminosity and other elements that cancel in this
ratio. This is the subject of an ongoing study which will lead to
values of the differential cross section as a function of the
$\omega$ polar angle~\cite{Karin2}.

\vspace{1pt}We are grateful to the personnel at The Svedberg
Laboratory for their support during the course of the experiment.
This work was supported by the European Community under the
``Structuring the European Research Area'' Specific Programme
Research Infrastructures Action (Hadron Physics, contract number
RII3-cT-204-506078), and by the Swedish Research Council.

\vspace{1pt}

%
%%%%%%%%%%%%%%%%%%%%%%%%%%%%%%%%%%%%%%%%%%%%%%%%%%%%%%%%%%%%%%%%%
%

\end{document}